\documentclass[prd,showpacs,showkeys,nofootinbib,floatfix,
               fleqn,preprint,12pt,tightenlines]{revtex4-1}

\usepackage{amsmath,amssymb,revsymb,graphicx,dcolumn}
\usepackage{array,hyperref,rotating}
\usepackage{dcolumn}  

\newcommand{\beq}{\begin{equation}}
\newcommand{\eeq}{\end{equation}}
\newcommand{\beqa}{\begin{eqnarray}}
\newcommand{\eeqa}{\end{eqnarray}}
\newcommand{\bsubeqs}{\begin{subequations}}
\newcommand{\esubeqs}{\end{subequations}}

\newcommand{\half}{{\textstyle \frac{1}{2}}}

\begin{document}

\begin{widetext}
\noindent Acta Phys. Pol. B \textbf{54}, 7-A3 (2023) \hfill  arXiv:2305.13278
%
%
\newline\vspace*{0mm}
\end{widetext}

\title{Vacuum-defect wormholes and a mirror world}

 \author{\vspace*{4mm}F.R. Klinkhamer}
 \email{frans.klinkhamer@kit.edu}
 \affiliation{Institute for Theoretical Physics,
 Karlsruhe Institute of Technology (KIT),\\
 76128 Karlsruhe, Germany\\}

\vspace*{10mm}

\begin{abstract}
\vspace*{2.5mm}\noindent
We have recently discovered a smooth vacuum-wormhole solution
of the first-order equations of general relativity.
Here, we obtain the corresponding multiple-vacuum-wormhole solution.
Assuming that our world is essentially Minkowski spacetime with
a large number of these vacuum-defect wormholes inserted, 
there is then another flat spacetime with opposite
spatial orientation, which may be called a ``mirror'' world.
We briefly discuss some phenomenological aspects and point out
that there will be no significant vacuum-Cherenkov radiation
in our world, so that ultrahigh-energy cosmic rays
do not constrain the typical size and separation of the  
wormhole mouths (different from the constraints obtained
for a single Minkowski spacetime with similar defects).
Other signatures from a ``gas'' of vacuum-defect wormholes 
are mentioned, including a possible time machine. 
\vspace*{10mm}
\end{abstract}


\maketitle

\section{Introduction}
\label{sec:Introduction}

Traversable wormholes appear to require some form of exotic input:
exotic  
matter~\cite{Ellis1973,Bronnikov1973,MorrisThorne1988,Visser1996,%
Rubakov2014,GaoJafferisWall2017,Kundu2022} 
or, as recently pointed out,
an exotic spacetime metric~\cite{Klinkhamer2023}.
Specifically, the new solution
is characterized by possessing a particular type of degenerate metric
(having a 3-dimensional hypersurface with vanishing determinant)
and does not require the presence of exotic matter.
Physically, the degenerate hypersurface corresponds to a
``spacetime defect,''
as discussed in previous papers~\cite{Klinkhamer2014,%
KlinkhamerSorba2014,Klinkhamer2018}.
(The origin of such a nonsmooth structure  
may be with the emergence of spacetime itself, 
as a kind of crystallization of ``spacetime atoms.''
An exploratory calculation relies on the IIB-matrix-model 
formulation~\cite{IKKT-1997,Aoki-etal-review-1999}
of nonperturbative superstring theory, where the emerging classical
spacetime is encoded in the so-called master 
field~\cite{Klinkhamer2021-master,Klinkhamer2021-regBB,%
Klinkhamer2022-corfu}.)

An explicit example of this new type of wormhole solution
is given by the vacuum-defect-wormhole solution,
which does not rely on any form of matter, exotic or not.
This vacuum-defect wormhole provides, in fact,
a \emph{smooth} solution of the first-order equations of
general relativity~\cite{Horowitz1991} and a brief summary
of this mathematical result will be given later on.

This last vacuum solution corresponds to a single wormhole
and the present paper discusses 
the generalization to multiple-vacuum-wormhole solutions.  
That construction is straightforward but the interpretation
is not.

If we live in a nearly flat world 
with many vacuum-defect wormholes inserted,  
then this is only possible if
there exists also a \emph{mirror world} with opposite spatial orientation.
The potential existence of
a mirror world (or mirror universe)  has, of course,
been discussed before; a selection of research papers is
given in Refs.~\cite{LeeYang1956,KobzarevOkunPomeranchuk1966,%
BlinnikovKhlopov1982,SenjanovicWilczekZee1984,Linde1988,Mohapatra-etal2002,%
Das-etal2011,Berezhiani-etal-EPJC-2018,Dunsky-etal2023}
and two review papers appear in Refs.~\cite{Okun2007,Foot2014}.
But the hypothetical mirror world we obtain 
from our vacuum-defect wormholes  
is rather different from the mirror worlds
discussed in the literature. We present, therefore, some exploratory
remarks on the vacuum-defect-wormholes phenomenology,
postponing a more detailed treatment
to the future. Throughout this paper, we use natural units
with $c=1$  and $\hbar=1$, unless stated otherwise.

\section{Single vacuum-defect wormhole}  
\label{sec:Single-vacuum-defect-wormhole}

\subsection{Tetrad and connection}
\label{subsec:Tetrad-connection}

Let us give a succinct description of the vacuum-defect-wormhole solution
(labeled ``vac-def-WH-sol'' below), 
with further details in Ref.~\cite{Klinkhamer2023}.
We will use the differential-form notation of
Ref.~\cite{EguchiGilkeyHanson1980}
and collect some basic equations of the first-order formulation
of general relativity~\cite{Schrodinger1950,Ferraris-etal-1982,%
Wald1984,Burton1998}
in Appendix~\ref{app:First-order-vacuum-equations-of-GR}.
Different from the standard second-order formulation of
general relativity, the first-order formulation
does not require the inverse metric $g^{\mu\nu}(x)$,
the metric $g_{\mu\nu}(x)$ [or, equivalently, the
tetrad $e^{a}_{\phantom{z}\mu}(x)$] suffices.

The spacetime coordinates are assumed to be given by
\bsubeqs\label{eq:single-wormhole-coordinates}
\beqa
t &\in& (-\infty,\,\infty)\,,
\eeqa       
\beqa       
\xi &\in& (-\infty,\,\infty)\,,
  \\[2mm]
\theta &\in& [0,\,\pi]\,,
\\[2mm]
\phi &\in& [0,\,2\pi)\,,
\eeqa
\esubeqs
where $\theta$ and $\phi$ are the standard spherical polar coordinates.
The proposed tetrad $e^{a}_{\phantom{z}\mu}(x)$ follows
from the following dual basis
$e^{a} \equiv e^{a}_{\phantom{z}\mu}\,\text{d}x^{\mu}$\,:
\bsubeqs\label{eq:vacuum-wormhole-tetrad}
\beqa
e^{0}\,\Big|_\text{vac-def-WH-sol} 
&=& \text{d}t\,,
\\[2mm]
\label{eq:vacuum-wormhole-tetrad-a-is-1}
e^{1}\,\Big|_\text{vac-def-WH-sol}  
&=&  \frac{\xi}{\sqrt{b^{2} + \xi^{2}}}\; \text{d}\xi\,,
\\[2mm]
e^{2}\,\Big|_\text{vac-def-WH-sol}   
&=&  \sqrt{b^{2} + \xi^{2}}\; \text{d}\theta\,,
\\[2mm]
e^{3}\,\Big|_\text{vac-def-WH-sol} 
&=& \sqrt{b^{2} + \xi^{2}}\;\sin\theta  \;  \text{d}\phi\,.
\eeqa
\esubeqs
The proposed connection $\omega^{\phantom{z}a}_{\mu\phantom{z}b}(x)$
has the following nonzero components of the corresponding 1-form:
\beqa
\label{eq:vacuum-wormhole-connection}
\left\{
\omega^{2}_{\phantom{z}1},\,
\omega^{3}_{\phantom{z}1},\,
\omega^{3}_{\phantom{z}2}
\right\}\,\Big|_\text{vac-def-WH-sol}   
&=&
\left\{
-\omega^{1}_{\phantom{z}2},\,
-\omega^{1}_{\phantom{z}3},\,
-\omega^{2}_{\phantom{z}3}
\right\}\,\Big|_\text{vac-def-WH-sol}  
\nonumber\\[2mm]
&=&
\left\{
\text{d}\theta,\,
\sin\theta  \;  \text{d}\phi,\,
\cos\theta  \;  \text{d}\phi
\right\}\,,
\eeqa

From this tetrad and connection, we obtain a vanishing curvature
2-form $R^{\,a}_{\,\phantom{z}b} \equiv \text{d}\omega^{\,a}_{\,\phantom{z}b}+
\omega^{a}_{\phantom{z}c}  \wedge \omega^{c}_{\phantom{z}b}\,$,
\beq
\label{eq:vacuum-wormhole-curvature}
R^{\,a}_{\phantom{z}b}\,\Big|_\text{vac-def-WH-sol}  
=0\,,
\eeq
which corresponds to Riemann tensor
$\mathcal{R}_{\kappa\lambda\mu\nu}(x)=0$
in the standard coordinate formulation
(the calligraphic symbol indicates the difference with
the curvature 2-form $R^{\,a}_{\,\phantom{z}b}$).
 
For later reference, the metric $g_{\mu\nu}(x) \equiv   
e^{a}_{\phantom{z}\mu}(x)\,e^{b}_{\phantom{z}\nu}(x)\,\eta_{ab}$ is given by  
the following line element:
\beq
\label{eq:vacuum-wormhole-metric}
ds^{2}\,\Big|_\text{vac-def-WH-sol}  
=
- dt^{2} + \frac{\xi^{2}}{b^{2} + \xi^{2}}\;d\xi^{2}
+ \left(b^{2} + \xi^{2}\right)\,
  \Big[ d\theta^{2} + \sin^{2}\theta\, d\phi^{2} \Big]\,.
\eeq
With the metric component $g_{11}(x)=\xi^{2}/(b^{2} + \xi^{2})$,
this metric is noninvertible at $\xi=0$.

The tetrad from \eqref{eq:vacuum-wormhole-tetrad}
and the connection from \eqref{eq:vacuum-wormhole-connection}
are perfectly smooth at $\xi=0$. They solve
the first-order vacuum equations of general relativity,
as given by \eqref{eq:first-order-eqs-no-torsion}
and \eqref{eq:first-order-eqs-Ricci-flat}
in Appendix~\ref{app:First-order-vacuum-equations-of-GR}.
Alternatively, we can verify that the metric
$g_{\mu\nu}(x)$ from \eqref{eq:vacuum-wormhole-metric}
solves the second-order vacuum equation of general relativity,
$\mathcal{R}_{\mu\nu}(x)=0$,
defined at $\xi=0$ by continuous extension from
its limit $\xi \to 0$  (see, in particular,
Section~3.3.1 of Ref.~\cite{Guenther2017}).

\subsection{Spatial orientability}
\label{subsec:Spatial-orientability}

As preparation for the construction of multiple wormholes,
it will be useful to define already two sets of Cartesian coordinates
(one for the ``lower'' world with $\widetilde{l} < -b$
and the other for the ``upper'' world with $\widetilde{l} > b$):
\bsubeqs\label{eq:Cartesian-coordinates}
\beqa
\label{eq:Cartesian-coordinates-plus}
\left\{
\begin{array}{c}
  Z_{+} \\
  Y_{+} \\
  X_{+}
\end{array}
 \right\}
&=& \widetilde{l}(\xi)\;
\left\{
\begin{array}{l}
  \phantom{Q}\hspace*{-3mm}\cos\theta\\
  \phantom{Q}\hspace*{-3mm}\sin\theta\,\sin\phi \\
  \phantom{Q}\hspace*{-3mm}\sin\theta\,\cos\phi
\end{array}
 \right\}\,, \;\;\;\;\text{for}\;\;
 \widetilde{l}(\xi)\equiv  \xi\;\sqrt{1+b^{2}/\xi^{2}} \geq b>0\,,
\\[2mm]
\label{eq:Cartesian-coordinates-minus}
\left\{
\begin{array}{c}
  Z_{-} \\
  Y_{-} \\
  X_{-}
\end{array}
 \right\}
&=& \widetilde{l}(\xi)\;
\left\{
\begin{array}{l}
  \phantom{Q}\hspace*{-3mm}\cos\theta\\
  \phantom{Q}\hspace*{-3mm}\sin\theta\,\sin\phi \\
  \phantom{Q}\hspace*{-3mm}\sin\theta\,\cos\phi
\end{array}
 \right\}\,, \;\;\;\;\text{for}\;\;
 \widetilde{l}(\xi)\equiv  \xi\;\sqrt{1+b^{2}/\xi^{2}}\leq -b<0\,,
\\[2mm]
\label{eq:Cartesian-coordinates-antipodal}
 \left\{Z_{+},\,  Y_{+},\,  X_{+}\right\}
&\stackrel{\wedge}{=}&
 \left\{Z_{-},\,  Y_{-},\,  X_{-}\right\}\,,
\hspace*{16mm}\text{for}\;\; |\,\widetilde{l}(\xi)\,| = b\,.
\eeqa
\esubeqs
The last equation implements the identification of
``antipodal'' points on the two 2-spheres $S^{\,2}_{\pm}$
with $\widetilde{l}=+b$ or $\widetilde{l}=-b$.

In terms of the $\{t,\, \widetilde{l},\, \theta,\,  \phi\}$
coordinates, we have the flat metric
\beq
\label{eq:ltilde-metric}
ds^{2}=
- dt^{2} + d\widetilde{l}^{\;2}
+ \widetilde{l}^{\;2}\,
  \Big[ d\theta^{2} + \sin^{2}\theta\, d\phi^{2} \Big]\,.
\eeq
For an arbitrary fixed time $t$,
this corresponds to two flat Euclidian 3-spaces
with two open balls of equal radius $b$
excised and ``antipodal'' identifications on the borders of the balls.
Still, these coordinates $\{t,\, \widetilde{l},\, \theta,\,  \phi\}$
are only useful \emph{outside} the wormhole throat
($\widetilde{l} > b$ or $\widetilde{l} < -b$)
and not for the \emph{whole} manifold
(including the wormhole throat at $\widetilde{l}=\pm b$) .
See further discussion in the first technical remark
of Section~III B in Ref.~\cite{Klinkhamer2023},
which contains additional references.  

Let us, finally, remark that the vacuum-defect-wormhole solution
of the present section is of the \emph{inter}-universe type,
connecting two \emph{distinct} asymptotically-flat spaces
(cf. Fig.~1a of Ref.~\cite{MorrisThorne1988} and Fig.~1.1
of Ref.~\cite{Visser1996} or, more schematically,
Fig.~1 of Ref.~\cite{Klinkhamer2023}),
whereas an \emph{intra}-universe wormhole
connects to a \emph{single} asymptotically-flat space
(cf. Fig.~1b of Ref.~\cite{MorrisThorne1988} and Fig.~1.2
of Ref.~\cite{Visser1996}).
Our two different 3-spaces have opposite orientations,
according to the sign flip in \eqref{eq:vacuum-wormhole-tetrad-a-is-1}
for positive and negative values of $\xi$
[and the different signs of $\widetilde{l}$
in \eqref{eq:Cartesian-coordinates-plus}
and \eqref{eq:Cartesian-coordinates-minus}].

The factor $\xi/(b^{2} + \xi^{2})^{1/2}$
in the tetrad component $e^{1}_{\phantom{z}\mu}(x)$
from \eqref{eq:vacuum-wormhole-tetrad-a-is-1}
is essential for obtaining a smooth solution.
A factor $[\xi^2/(b^{2} + \xi^{2})]^{1/2}$
would give, from the
no-torsion condition \eqref{eq:first-order-eqs-no-torsion},
singular terms $\xi/(\xi^{2})^{1/2}$
in certain connection components
$\omega^{\phantom{z}a}_{\mu\phantom{z}b}(x)$.
In this way, we see that the change in spatial orientability
of the two asymptotically-flat spaces is a direct
consequence of having a smooth solution of
the first-order field equations of general relativity.

\section{Multiple vacuum-defect wormholes}  
\label{sec:Multiple-vacuum-defect-wormholes}

\subsection{Construction}
\label{subsec:Construction}

For the description of the multiple-vacuum-defect-wormhole solution,  
we need two definitions. First, there is
the embedding space $M_\text{embed}$,
which consists of the union of two copies of
Euclidean 3-space $E_{3}^{\,(\pm)}$,
one copy being labeled by `$+$' (the ``upper'' world)
and the other by `$-$' (the ``lower'' world).
Each of these 3-spaces has standard Cartesian coordinates,
so that we have
\bsubeqs
\beqa
\label{eq:embedding-space-def}
M_\text{embed} &=&
E_{3}^{\,(+)} \cup  E_{3}^{\,(-)}
\\[2mm]
\label{eq:embedding-space-coordinates}
E_{3}^{\,(\pm)} &:&
\left(X_{\pm},\,  Y_{\pm},\,  Z_{\pm}\right) \;\in\; \mathbb{R}^{3}\,.
\eeqa
\esubeqs
The multiple-vacuum-defect-wormhole solutions only cover
part of $M_\text{embed}$.

Second, we introduce the following definitions (wormhole label
$n \in \{1,\, 2,\, 3,\, \ldots\,,\,  N\}$):
\bsubeqs\label{eq:Cartesian-coordinates-n}
\beqa\label{eq:Cartesian-coordinates-n-ZYX}
\left\{
\begin{array}{c}
  Z_{\pm}^{(n)}\\
  Y_{\pm}^{(n)} \\
  X_{\pm}^{(n)}
\end{array}
 \right\}
&\equiv& \widetilde{l}_{n}\left(\xi_{n}\right)\;
\left\{
\begin{array}{l}
  \phantom{Q}\hspace*{-3mm}\cos\theta_{n}\\
  \phantom{Q}\hspace*{-3mm}\sin\theta_{n}\,\sin\phi_{n} \\
  \phantom{Q}\hspace*{-3mm}\sin\theta_{n}\,\cos\phi_{n}
\end{array}
 \right\}\,,
\\[2mm]
\widetilde{l}_{n}\left(\xi_{n}\right) &\equiv&
 \xi_{n}\;\sqrt{1+b_{n}^{2}/\xi_{n}^{2}}
\;\in\; (-\infty,\,-b_{n}] \cup [b_{n},\,\infty)
 \,,
\\[2mm]
\xi_{n} &\in& (-\infty,\,\infty)\,,
\quad
\theta_{n} \;\in\; [0,\,\pi]\,,
\quad
\phi_{n} \;\in\; [0,\,2\pi) \,,
\eeqa
\esubeqs
where the suffix `$+$' on the left-hand side of
\eqref{eq:Cartesian-coordinates-n-ZYX} holds for $\xi_{n}\geq 0$
and the suffix `$-$' for $\xi_{n}\leq 0$,
with ``antipodal'' identifications at $\xi_{n}= 0$ .
Note that we allow for wormholes of different sizes $b_{n}>0$.
The typical size of the wormholes will be denoted by $\overline{b}$.

We now give the explicit construction of the $N=2$
multiple-vacuum-defect-wormhole solution,
with an obvious generalization to larger values of $N$.
For the first wormhole ($n=1$) of the pair,
we have the following coordinates in $M_\text{embed}$:
\bsubeqs
\beqa\label{eq:N2-vac-def-WH-n-is-1}
\left\{
\begin{array}{c}
  Z_{\pm}\\
  Y_{\pm}\\
  X_{\pm}
\end{array}
 \right\}_{N=2}
&=&
\left\{
\begin{array}{c}
  Z_{\pm}^{(1)}\\
  Y_{\pm}^{(1)} \\
  X_{\pm}^{(1)}
\end{array}
 \right\}\,,
\,\;\;\;\;\;\;\;\;\text{for}\;\;X_{\pm} \leq l_{12}/2\,,
\eeqa
where the right-hand-side entries are defined in
\eqref{eq:Cartesian-coordinates-n}
and where $l_{12}>0$ will be given shortly.
This first wormhole is centered at the spatial origins of both worlds,
$(X_{\pm},\, Y_{\pm},\,Z_{\pm})
=\big(\widehat{X}_{1},\, \widehat{Y}_{1},\,\widehat{Z}_{1}\big) =(0,\,0,\,0)$,
and has open balls of equal radius $b_{1}$ removed from
the two flat spaces.

\begin{figure}[p]   
\vspace*{-2mm}
\begin{center}   
\includegraphics[width=0.5\textwidth]{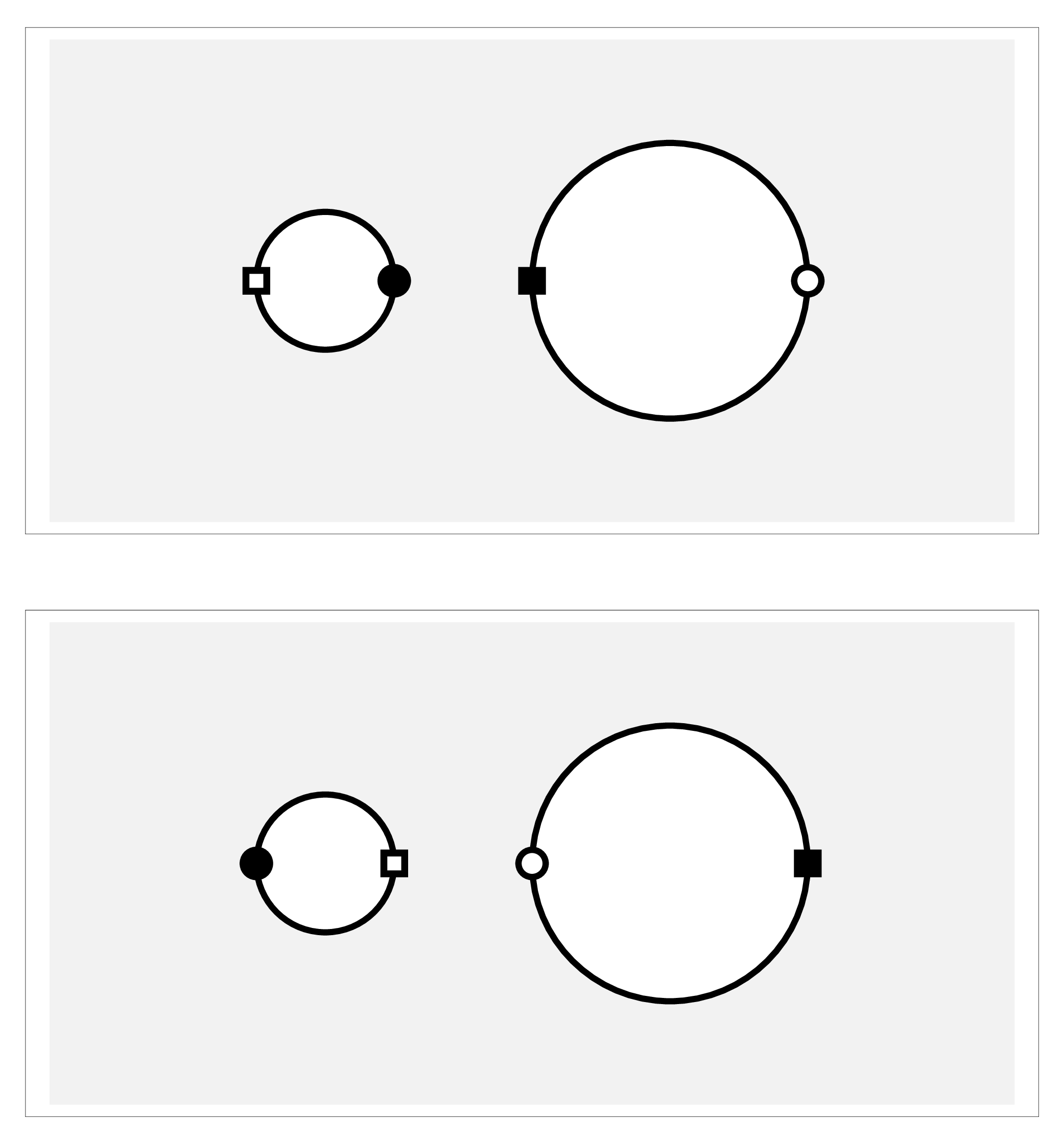}
\end{center}
\vspace*{-7mm}
\caption{Sketch of the $N=2$
multiple-vacuum-defect-wormhole spacetime,
at $Z_{\pm}=0$ and an arbitrary fixed time $t$.
The top panel in this figure corresponds to the
``upper'' world and the bottom panel to the ``lower'' world.
The two worlds are connected by vacuum-defect wormholes.
Here, two wormhole throats
are shown as heavy circles with ``antipodal'' spacetime points identified
(four distinct spacetime points on the two wormhole mouths
are marked by different symbols). 
}
\label{fig:N-is-2-multiple-wormhole-points}
\vspace*{00mm}
\vspace*{1mm}
\vspace*{-0mm}
\begin{center}   
\includegraphics[width=0.5\textwidth]{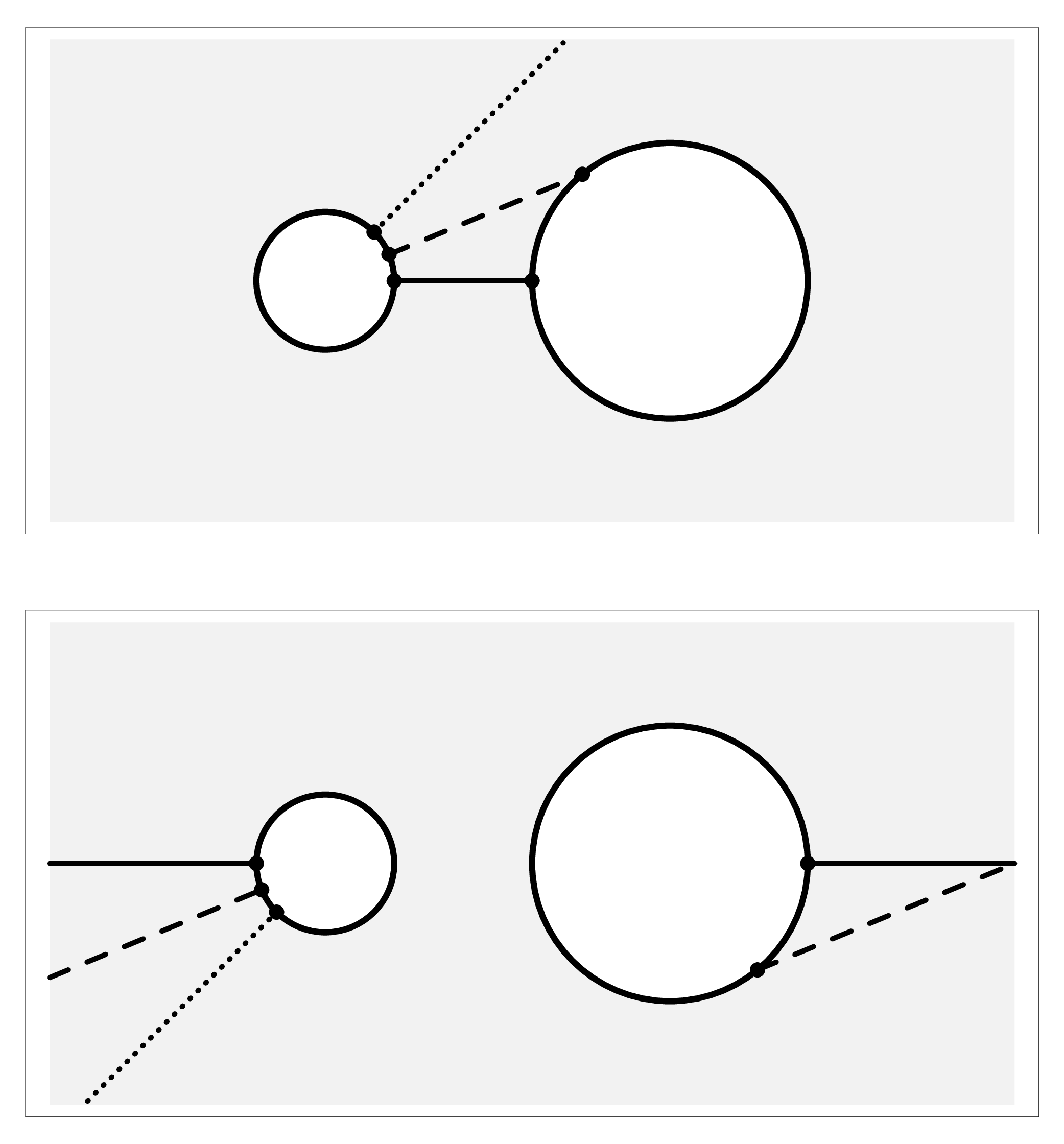}
\end{center}
\vspace*{-7mm}
\caption{Paths of three light rays in the
$N=2$ multiple-vacuum-defect-wormhole spacetime
from Fig.~\ref{fig:N-is-2-multiple-wormhole-points}.
The light rays start in the ``lower'' world on the left
and the small dots on the wormhole throats are purely indicative.
}
\label{fig:N-is-2-multiple-wormhole-light-rays}
\vspace*{00mm}
\end{figure}

For the second wormhole ($n=2$) of the pair,
we only discuss a simple case,
namely an equal translation along the $X_{\pm}$ axes
in both worlds:
\beqa\label{eq:N2-vac-def-WH-n-is-2}
\left\{
\begin{array}{c}
  Z_{\pm}\\
  Y_{\pm}\\
  X_{\pm}
\end{array}
 \right\}_{N=2}
&=&
\left\{
\begin{array}{c}
  Z_{\pm}^{(2)}\\
  Y_{\pm}^{(2)} \\
  X_{\pm}^{(2)} + l_{12}
\end{array}
 \right\}\,,
\,\;\;\;\;\;\;\;\;\text{for}\;\;X_{\pm} \geq l_{12}/2\,,
\eeqa
with
\beqa\label{eq:N2-vac-def-WH-d12}
l_{12}\;\Big|_{N=2} &>& b_{1}+b_{2}\,.
\eeqa
\esubeqs
This second wormhole is centered on
$(X_{\pm},\, Y_{\pm},\,Z_{\pm})
=\big(\widehat{X}_{2},\, \widehat{Y}_{2},\,\widehat{Z}_{2}\big)
=(l_{12},\,0,\,0)$ in  both worlds
and has open balls of equal radius $b_{2}$ removed from
the two flat spaces.  A sketch appears in
Fig.~\ref{fig:N-is-2-multiple-wormhole-points}.
The two balls in the two copies of $\mathbb{R}^3$
do not overlap or touch, provided their separation is large
enough, as guaranteed by condition \eqref{eq:N2-vac-def-WH-d12}.
The $N=2$ solution parameters are thus the wormhole radii $b_{n}$
and the  wormhole
centers $\big(\widehat{X}_{n},\, \widehat{Y}_{n},\,\widehat{Z}_{n}\big)$,
for $n=1$ and $n=2$.

To get  $N\geq 3$ multiple vacuum-defect wormholes,
we need to add more single wormholes,
appropriately shifted so that their excised balls do not overlap or
touch. The typical separation of the wormhole mouths
will be denoted by $\overline{l}$ and is defined by the wormhole
number density $\overline{n}\equiv 1/\overline{l}^{\,3}$.

\subsection{Metric}
\label{subsec:Metric}

We can be relatively brief about the (flat) metric
of the multiple-vacuum-defect-wormhole solution
(labeled ``multiple-vac-def-WH-sol'' below). 
In between the wormhole mouths, there is the standard
flat metric in either 3-space (marked $\pm$),  
\beq
\label{eq:outside-metric}
ds^{2}\,\Big|_\text{multiple-vac-def-WH-sol}^{\text{(outside\;WH-mouths)}}
=
- dt^{2} + \left(d X_{\pm} \right)^{2}
+ \left(d Y_{\pm} \right)^{2}
+ \left(d Z_{\pm} \right)^{2}\,,
\eeq
for the spatial coordinates \eqref{eq:embedding-space-coordinates}
of the embedding space.

For the metric at or near the wormhole mouths,
we have essentially the metric \eqref{eq:vacuum-wormhole-metric}.
Considering, for example, the particular wormhole mouth
with label $\overline{n}$, the metric is
\beq
\label{eq:1}
ds^{2}\,\Big|_\text{multiple-vac-def-WH-sol}^{\text{(near WH-mouth $\overline{n}$)}}
=
- dt^{2}
+ \frac{\xi_{\,\overline{n}}^{2}}{b_{\,\overline{n}}^{2}
+ \xi_{\,\overline{n}}^{2}}\;d\xi_{\,\overline{n}}^{2}
+ \left(b_{\,\overline{n}}^{2} + \xi_{\,\overline{n}}^{2}\right)\,
  \Big[ d\theta_{\,\overline{n}}^{2}
        + \sin^{2}\theta_{\,\overline{n}}\, d\phi_{\,\overline{n}}^{2} \Big]\,,
\eeq
with coordinates
$\theta_{\,\overline{n}} \in [0,\,\pi]$,
$\phi_{\,\overline{n}} \in [0,\,2\pi)$, and
$\xi_{\,\overline{n}} \in [-\Delta_{\,\overline{n}},\,\Delta_{\,\overline{n}}]$
for a positive infinitesimal $\Delta_{\,\overline{n}}\,$.
The corresponding tetrad is given by \eqref{eq:vacuum-wormhole-tetrad},
but now in terms of the spatial coordinates
$\{\xi_{\,\overline{n}},\,\theta_{\,\overline{n}},\, \phi_{\,\overline{n}}\}$.
The other $N-1$ wormholes have, near their mouths,
tetrads of \emph{identical} structure, which makes for a
\emph{consistent} spatial orientation of the
``upper'' world, as well as of the ``lower'' world.

\section{Phenomenology}
\label{sec:Phenomenology}

\subsection{Preliminary remarks}
\label{subsec:Preliminary-remarks}

Let us explore some of the phenomenology
from a flat (but nontrivial) spacetime corresponding
to a multiple-vacuum-defect-wormhole solution. Physically, we
start from an empty spacetime with a ``gas'' of randomly sprinkled
static vacuum-defect wormholes and a small number of material test particles.
Some of the details of the construction will become clear as we
progress in the present section.

As to cosmology, we are completely agnostic and leave that
discussion to
a future paper. The problem is that the cosmology of our two-sheeted
spacetime requires many further assumptions,
for example, if ``our'' world
has a matter-antimatter asymmetry as observed,
what about the other world?

\subsection{Light rays}
\label{subsec:Light-rays}

Consider light rays in an empty spacetime 
with vacuum-defect wormholes, 
where the flat metric between the wormhole mouths
is given by \eqref{eq:outside-metric}.
The case of a single wormhole was already discussed in Appendix~A of
Ref.~\cite{Klinkhamer2023}.
What happens for the case of more wormholes
is already clear from the $N=2$ solution.

Figure~\ref{fig:N-is-2-multiple-wormhole-light-rays}
shows three light rays starting out at the left of the
``lower'' world and reaching the throat of the left wormhole.
For the special light ray along the $N=2$ wormhole-displacement vector
(solid curves in Fig.~\ref{fig:N-is-2-multiple-wormhole-light-rays}),
the light ultimately continues towards the right
of the ``lower'' world, after a brief detour in the ``upper'' world.
A similar behavior holds for light rays close to this special ray,
except that the returned rays in the ``lower'' world are parallel
shifted (see the in- and outgoing dashed curves in the
bottom panel of Fig.~\ref{fig:N-is-2-multiple-wormhole-light-rays}).
But for light rays sufficiently
far from the initial special ray, the rays get lost in the
 ``upper'' world, as shown by the dotted curves
in Fig.~\ref{fig:N-is-2-multiple-wormhole-light-rays}.

Let us continue the discussion of that last light ray (dotted curves
in Fig.~\ref{fig:N-is-2-multiple-wormhole-light-rays})
for the case of a dense gas of wormholes. The dotted curve
in the ``upper'' world of
Fig.~\ref{fig:N-is-2-multiple-wormhole-light-rays}
will then ultimately hit a wormhole and return
to ``lower'' world.
In this way, the propagation of light rays in the ``lower'' world
gets modified by the presence of vacuum-defect wormholes, 
with a random-walk of parallel shifts (leading to a
blurred image of a point source).

\subsection{Dispersion relations}
\label{subsec:Dispersion-relations}

Several years ago, we have calculated~\cite{BernadotteKlinkhamer2006}
the modifications of the photon dispersion relation
due to a Swiss-cheese-type spacetime
with a ``gas'' of randomly-positioned static spacetime defects.
The results for the modified dispersion relations of photons
and Dirac particles are summarized in
Appendix~\ref{app:Modified dispersion-relations-from-spacetime-defects}.
The calculation outlined in that appendix
was for a ``gas'' of static defect in a \emph{single} flat spacetime.
The question now is
what happens in our hypothetical \emph{two-sheeted} world
(as sketched in Figs.~\ref{fig:N-is-2-multiple-wormhole-points}
and \ref{fig:N-is-2-multiple-wormhole-light-rays}).

Recall that, for photons in the single world with localized
defects, we are after the solution of the vacuum
Maxwell equations in a flat spacetime but with special boundary conditions
from the defects
(see Fig.~\ref{fig:N-is-2-multiple-defect-points}  in
Appendix~\ref{app:Modified dispersion-relations-from-spacetime-defects}).
The effects of these boundary conditions are described by fictitious
multipoles located inside the holes (this particular construction
goes back to Bethe in 1944, who applied it to wave guides).
The photon dispersion relation is modified by the fields of
these multipoles; see Section~II B of Ref.~\cite{BernadotteKlinkhamer2006},
with further details in Chaps.~3 and 4 of Ref.~\cite{Bernadotte2006}
and Chaps.~12 and 13 of Ref.~\cite{Schwarz2010}.

But if we now turn to our two-sheeted world
and  first look at what happens in the lower
world, then we see that there
are \emph{no} special boundary conditions on the fields
at the wormhole mouths,
as shown in the bottom panel of
Fig.~\ref{fig:N-is-2-multiple-wormhole-points}
(whereas there \emph{are} special boundary conditions on the fields
at the defect locations in Fig.~\ref{fig:N-is-2-multiple-defect-points}).
Hence, as far as the lower world is concerned, we have the
standard plane wave solutions of the Maxwell and Dirac equations
with standard dispersion relations.

Still, there are special boundary conditions, but they
connect the fields of the lower world to those in the upper world
(after a parity transformation, in fact).
So, if we have a standard plane wave in the lower world,
then we have essentially the same standard plane wave in the upper world
but with reversed 3-momentum, $\vec{k}_{U}=-\vec{k}_{L}$.
For these unmodified plane waves (labeled ``$\text{LU-sol}$''), 
the dispersion relations
of photons ($\gamma$) and structureless Dirac particles ($p$)
are simply
\bsubeqs\label{eq:mod-disp-rel-standard-LU-sol}
\beqa
\label{eq:photon-mod-disp-rel-standard-LU-sol}
\left[\omega_{\gamma}^\text{\,LU-sol}(k)\right]^2
&\sim&
c^2\,k^2 +  \ldots \,,
\\[2mm]
\label{eq:proton-mod-disp-rel-standard-LU-sol}
[\omega_{p}^\text{\,LU-sol}(k)]^2
&\sim&
c^2/\lambda_{p}^2 + c^2\,k^2  +  \ldots \,,
\eeqa
\esubeqs
with $k\equiv  |\vec{k}|$ and the reduced Compton wavelength
$\lambda_{p}\equiv \hbar/(m_{p}\,c)$ of the spin-$\half$
particle.
The ellipses in the above dispersion relations allow for further terms
appearing as corrections to the leading
large-wavelength ($k\,\overline{b} \ll 1$) and
dilute-gas ($\overline{b}/\overline{l} \ll 1$) approximations.

Anyway, it is clear from \eqref{eq:mod-disp-rel-standard-LU-sol}
that we do not expect significant
vacuum-Cherenkov radiation~\cite{KlinkhamerSchreck2008}, so that
ultrahigh-energy cosmic rays (UHECRs) do not constrain
the values of the typical wormhole size $\overline{b}$
and the typical wormhole separation $\overline{l}$
(different from the UHECR bounds on defect length scales as reviewed
in Appendix~\ref{app:Modified dispersion-relations-from-spacetime-defects}).

\subsection{Scattering}
\label{subsec:Scattering}

Dispersion relations, modified or not,
correspond to essentially stationary
phenomena. But we can also use transient phenomena
to probe the presence of vacuum-defect wormholes.

Consider a source emitting a light pulse
and a distant observer, both located in our (``lower'') world.
Then, the wormhole mouths act like perfect absorbers (the light
is gone from the lower world).
But over a long time the light is returned to the lower
world by other wormholes (cf. the solid and dashed curves
in Fig.~\ref{fig:N-is-2-multiple-wormhole-light-rays}
for a single wormhole pair).
Averaging over time, the wormhole mouths in the lower world
both absorb and emit.

The actual magnitude of scattering effects from the
wormhole mouths would be expected to be reduced compared
to that of the case of defects, as calculated in Section~II C of
Ref.~\cite{BernadotteKlinkhamer2006}, because the
wormhole case does not require fictitious dipoles
from the boundary conditions. Still, the wormhole
boundary conditions will somehow contribute to the scattering.
It remains to establish the scattering length $L_\text{scatt}$
and its dependence on $\overline{b}$ and $\overline{l}$,
possibly with some $k\,\overline{b}$ dependence at
smaller wavelengths.

Awaiting the definitive calculation of the scattering length,
we can already present a rough estimate.
Referring to the discussion in Section~II C of
Ref.~\cite{BernadotteKlinkhamer2006},
we take the effective wormhole cross section
$\sigma \sim \mathcal{F}\,\pi\,\overline{b}^{\,2}$
[with factor $\mathcal{F}=1$ for a black disk
and $\mathcal{F}<1$ from additional emission effects]
and the coherence number $N_\text{coh} \sim (k\,\overline{l}\,)^{-3}$
for the case considered, $\overline{b}\ll \overline{l}\ll \lambda$.
Then the absorption coefficient $\alpha$
(or the inverse of the scattering length $L$) reads
\beq
\label{eq:alpha-scatt}
\alpha_\text{scatt}^\text{(WH)}
\equiv 1/L_\text{scatt}^\text{(WH)}
\sim
\sigma \;\overline{l}^{\,-3} \;  N_\text{coh}
\sim
\mathcal{F}\,\pi\,\overline{b}^{\,2}\; k^{-3}\;\overline{l}^{\,-6}\,.
\eeq
We now demand that $L_\text{scatt}^\text{(WH)}$
be larger than the source distance $D$.

For some ballpark numbers, we can use the same $2\,\text{TeV}$
gamma-ray flare from Markarian 421
(distance $D \approx 3.8 \times 10^{24}\,\text{m}$)
as used in Sections~IV A and B of Ref.~\cite{BernadotteKlinkhamer2006}. 
Demanding $L_\text{scatt}^\text{(WH)} \gtrsim D/f$ for a factor $f > 1$,
we get
\beqa
\label{eq:scatt-bound}
\hspace*{-5mm}
\left(\,\overline{l}/\overline{b}\,\right)^{2}\;\overline{l}^{\,4}\;
\Big|^\text{(scatt)}
&\gtrsim&
\pi\,\mathcal{F}\;k^{-3}\;D/f
\nonumber \\[2mm]
\hspace*{-5mm}
&=&
1.15 \times 10^{-34}\,\text{m}^{4}\;
\left(\frac{\mathcal{F}}{1}\right)\,
\left(\frac{2.0\,\text{TeV}}{E_{\gamma}}\right)^{3}\,
\left(\frac{D}{3.8 \times 10^{24}\,\text{m}}\right)\,
\left(\frac{10^{2}}{f}\right)\,,
\eeqa
where the factor $f$ has been extensively discussed
in Section~IV B of Ref.~\cite{BernadotteKlinkhamer2006}
and where the actual value of the factor $\mathcal{F} \geq 0$
in the effective wormhole cross section
would follow from the definitive calculation.

For a spacetime with single-scale vacuum-defect wormholes 
($\overline{b}\sim \overline{l}\,$)
and with the assumption $\overline{l}\ll \lambda$,
bound \eqref{eq:scatt-bound} for the
$\{E_{\gamma},\,  D,\,f \}$ values stated gives
$\mathcal{F}^{\,1/4}\,10^{-8.5}\,\text{m}  \lesssim  \overline{l}
\ll 10^{-19}\,\text{m}$. Similarly,
for Planck-size wormholes ($\,\overline{b}=10^{-35}\,\text{m}$),
bound \eqref{eq:scatt-bound} gives
$\mathcal{F}^{\,1/6}\,10^{-17.3}\,\text{m}  \lesssim  \overline{l}\ll 10^{-19}\,\text{m}$.
Both examples suggest small values of $\mathcal{F}$.
But, as said, these numbers are only indicative.

\subsection{Imaging bound}
\label{subsec:Imaging-bound}

Following-up on the last remark of Section~\ref{subsec:Light-rays},
we can use high-resolution
imaging by optical microscopes to get bounds on the
length-scales of a ``gas'' of vacuum-defect wormholes 
(typical size $\overline{b}$  and typical separation $\overline{l}\,$).

For light travelling a distance $L$, there are
$N \sim L/\overline{l}$ encounters with the wormhole mouths.
As explained in Section~\ref{subsec:Light-rays},
the built-up parallel shift of the light ray
results from a random-walk process
and its order of magnitude is given by
\beq
\label{eq:Delta-x}
[\Delta x]_\text{random-walk} \sim \sqrt{N}\,\overline{b}
\sim \left(\,\overline{b}/\overline{l}\,\right)\;
\sqrt{L\,\overline{l}}\,.
\eeq
In order to get a clear image of an object with substructure
$[\delta x]_\text{object}$, we must have negligible random-walk
parallel shifts of the light ray,
\beq
\label{eq:clear-image-cond}
[\Delta x]_\text{random-walk} \lesssim [\delta x]_\text{object}\,.
\eeq
Putting in some (optimistic) numbers for a high-resolution
optical microscope~\cite{Mertz2019},
we get the following bound
on the vacuum-defect-wormhole length scales:
\beq
\label{eq:upper-bound-microscope}
\left(\,\overline{b}/\overline{l}\,\right)\,\overline{b}\;   
\Big|^\text{(opt.\;microscope)}
\lesssim
\frac{\left([\delta x]_\text{object}\right)^{2}}{L}
=
4 \times 10^{-13}\,\text{m}\;
\left(\frac{[\delta x]_\text{object}}{200\,\text{nm}}\right)^{2}\,
\left(\frac{0.1\,\text{m}}{L}\right)\,,
\eeq
where the actual $L$ value to be used depends on the details
of the instrument. 
The optical-microscope bound shown in \eqref{eq:upper-bound-microscope}  
is purely indicative.

Switching over to a transmission electron microscope
(TEM, spatial resolution of about $0.1\,\text{nm}= 1\,\text{\r{A}}$;
see, e.g., Ref.~\cite{Spence2017} for background and
Refs.~\cite{CreweWallLangmore-1970,Chen-etal-2021} for some further images), 
we have the potential to reduce
the upper bound \eqref{eq:upper-bound-microscope}
by several orders of magnitude.
For a dilute gas of vacuum-defect wormholes,  
clear TEM images would then give
\beq
\label{eq:upper-bound-TEM}
\left(\,\overline{b}/\overline{l}\,\right)\,\overline{b}\;
\Big|^\text{(TEM)}
\lesssim
10^{-18}\,\text{m}\;
\left(\frac{[\delta x]_\text{object}}{1\,\text{\r{A}}}\right)^{2}\,
\left(\frac{0.01\,\text{m}}{L}\right)\,,
\eeq
where \eqref{eq:upper-bound-microscope} has been used
with a modest value $L \sim 1\,\text{cm}$
(consistent with lens separations of order $10\,\text{cm}$
as given in Table~2.1 of Ref.~\cite{Spence2017}).
Ideally,
a special-purpose TEM would have condenser lenses designed
to produce more or less parallel rays after the rays
pass through a 2D sample
with sub\r{a}ngstr\"{o}m structure and to have these parallel
rays travelling
freely for about $L \sim 10\,\text{cm}$ (possibly broadened
by vacuum-defect-wormhole effects) before they enter the rest
of the microscope. In that case, the bound \eqref{eq:upper-bound-TEM}
could drop to a value of order $10^{-21}\,\text{m}$.

The electron-microscope
bound shown in \eqref{eq:upper-bound-TEM}
is, for the moment,  purely indicative.
Still, from \eqref{eq:upper-bound-TEM}
for the $[\delta x]_\text{object}$ and $L$ values stated,
it appears perfectly possible to have a gas of
Planck-scale vacuum-defect wormholes,  
$\,\overline{b} \sim \overline{l} \sim 10^{-35}\,\text{m}$.

Returning to light beams and bound \eqref{eq:upper-bound-microscope},
it is perhaps feasible to design an interferometer experiment
(with variable arm-lengths and whatever may be needed)
to test for random-walk broadening of the beams.  Possibly,
the LIGO expertise~\cite{LIGOScientific2007,LIGOScientific2014}
can be used for obtaining tight constraints on
$\left(\,\overline{b}/\overline{l}\,\right)\,\overline{b}$
(optimistically of order $10^{-41}\,\text{m}$, with
$[\delta x]_\text{object}$ replaced by $[\delta x]_\text{broadening}
\sim 10^{-19}\,\text{m}$ and $L \sim 1\,\text{km}$).

\subsection{Time machine}
\label{subsec:Time-machine}

It was realized already in the
original Morris--Thorne paper~\cite{MorrisThorne1988} that
traversable exotic-matter wormholes appear to
allow for backward time travel
(a simplified argument was presented in a follow-up
paper~\cite{MorrisThorneYurtsever1988}).
An especially elegant method for getting a time machine
was suggested by Frolov and Novikov~\cite{FrolovNovikov1990},
namely by putting a large mass near one of the mouths of an
intra-universe traversable wormhole, so that a clock near
that mouth runs slower (gravitational redshift)
than a clock near the other mouth.
The two clocks are then running at different rates
and they will get more and more out of step with each other.
A time machine appears when the two clocks are sufficiently
far out of step with each other.
(The previous two sentences paraphrase two sentences in
Section~18.3 of Ref.~\cite{Visser1996}; in fact,
Sections~18.1 and 18.3 of that reference
give a nice discussion of how chronology violation
appears to be inescapable if there are traversable wormholes.)

\begin{figure}[t]   
\vspace*{-0mm}
\begin{center}
\includegraphics[width=0.90\textwidth]{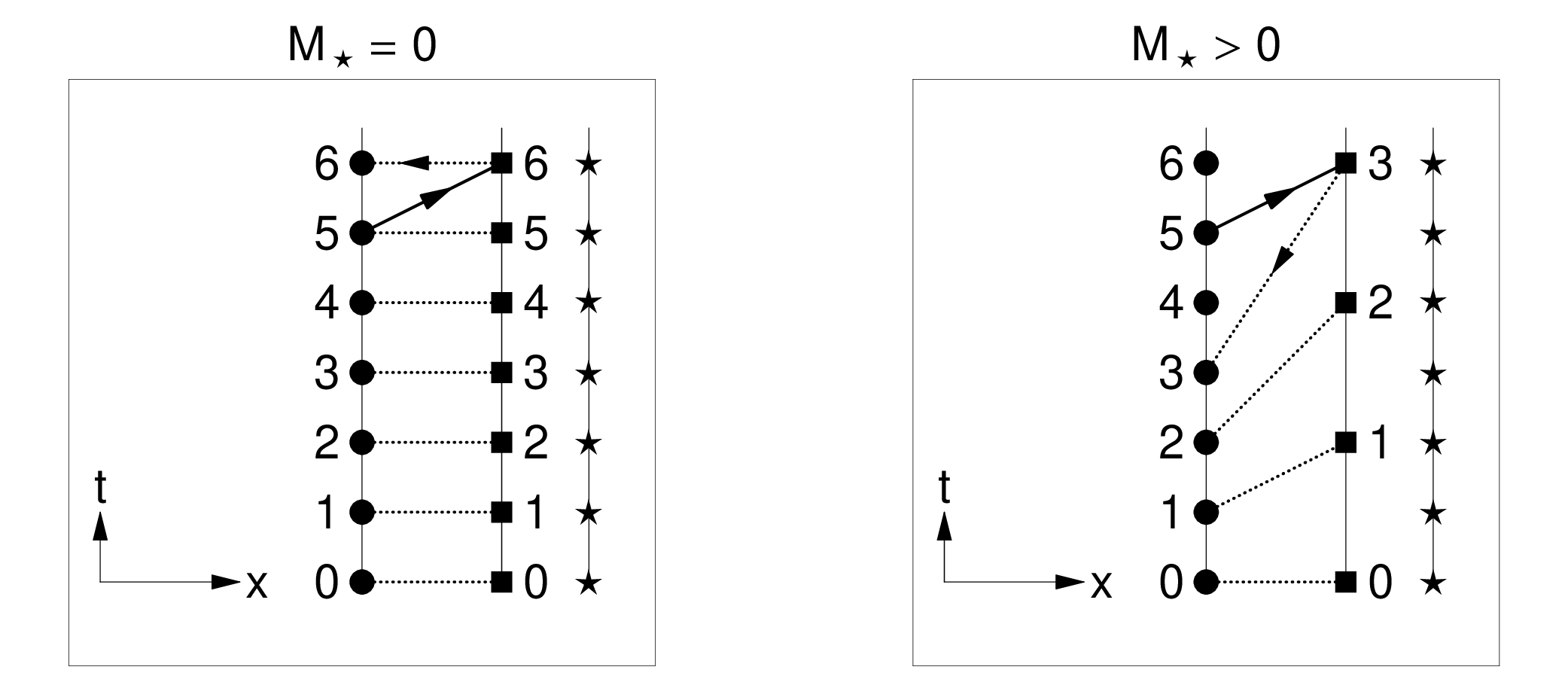}
\end{center}
\vspace*{-0mm}
\caption{Spacetime-diagram sketches (in rescaled units)
of two identical lower-world clocks near two wormhole mouths
(corresponding to the heavy dot and the
filled square in the bottom panel of
Fig.~\ref{fig:N-is-2-multiple-wormhole-points}).
At a lower-world  point close to the right wormhole mouth,
there can be a static point mass or not
(indicated by a star with $M_{\star}>0$ or $M_{\star}=0$).
Without a static point mass present (left panel), both clocks run at the
same rate: here, the ticks are shown for
$t_\text{left}=t_\text{right}=0,\,1,\,\ldots \,,\,6$.
A courageous explorer (with near-light speed)
starts out at local time
$t_\text{left}=5$ from a point near the left wormhole mouth
(heavy dot) on a long path in the lower world
(full line with arrow in the spacetime diagram of the left panel)
to reach a point near the right wormhole mouth (filled square)
and then passes quickly through the two wormhole throats
(dotted line with arrow in the spacetime diagram of the left panel)
back to the starting point (heavy dot) at local time $t_\text{left}=6$.
With a static point mass present (right panel), the clock
on the right runs slower than the clock on the left.
Again, an explorer starts out at local time
$t_\text{left}=5$ from a point near the left wormhole mouth
(heavy dot) on a long path in the lower world
(full line with arrow in the spacetime diagram of the right panel)
to reach
a point near the right wormhole mouth (filled square)
and then passes quickly through the two wormhole throats
(dotted line with arrow in the spacetime diagram of the right panel)
back to
the starting point (heavy dot) at local time $t_\text{left}=3$.
The explorer has travelled back in time
($t_\text{left,\,return}=3 < t_\text{left,\,start}=5$)
and we have effectively a time machine.
The time machine starts working for $t_\text{left} \geq 1$
(indeed, the same explorer starting out at local time
$t_\text{left}=1$ returns at local time $t_\text{left}=1$).
}
\label{fig:N-is-2-time-machine}
\end{figure}

It is not difficult to see that we can make a similar
time machine by use of multiple vacuum-defect wormholes
and a single localized mass. We proceed in two steps.

First, we note that a \emph{pair} of vacuum-defect wormholes
can effectively act as a \emph{single} intra-universe wormhole.
Consider, in fact, a point in the lower world near
the heavy dot of the bottom panel in
Fig.~\ref{fig:N-is-2-multiple-wormhole-points}.
While staying in the lower world, it is possible
to travel along a long path (a large semicircle, for example)
to a point near the filled square on the right.
Alternatively, it is possible in the lower world to enter
the left wormhole at the heavy dot,
to emerge in the upper world at the heavy dot,
to travel along a straight line in the upper world
towards the filled square,
to enter the right wormhole at the filled square,
and, finally, to re-emerge in the lower world at the filled square.
This last route can be very short if the two wormholes of the pair
are nearly touching ($l_{12}=b_{1} + b_{2} + \Delta l$ for
a positive infinitesimal $\Delta l$, in the notation of
Section~\ref{subsec:Construction}).

Second, we place a static point mass in the lower world just to the
right of the filled square of the bottom panel in
Fig.~\ref{fig:N-is-2-multiple-wormhole-points}. Then, a
lower-world clock near the filled square runs slower
(gravitational redshift)
than a lower-world clock near the heavy dot
and a time machine appears after a sufficiently long time
(see Fig.~\ref{fig:N-is-2-time-machine} with further details
in the caption).

Other constructions are certainly possible, but this simple example
suffices to show that time machines can, in principle,
appear if there exists, at least, one pair of vacuum-defect wormholes
and a single point mass which can be freely positioned
in one of the worlds.
The outstanding question concerns, as emphasized in
Ref.~\cite{FrolovNovikov1990},
the classical and quantum stability of this particular
type of time machine.

Still, it is truly remarkable that 
general relativity in the classical vacuum
(in the first-order formalism) 
appears to allow for backward time travel
via the existence of multiple vacuum-defect wormholes.
This is all the more surprising 
as recent results~\cite{GaoJafferisWall2017,Kundu2022}
on nonlocal matter quantum effects in a smooth classical spacetime  
suggest the existence of a traversable wormhole, 
but without the possibility of getting a time machine. Our time machine 
(Fig.~\ref{fig:N-is-2-time-machine})
relies on a nonsmooth spacetime structure, which, as mentioned in
Section~\ref{sec:Introduction}, 
may result from an early phase in the Universe, where spacetime itself 
(with or without defects) is 
created~\cite{IKKT-1997,Aoki-etal-review-1999,%
Klinkhamer2021-master,Klinkhamer2021-regBB,Klinkhamer2022-corfu}.

\section{Conclusion}\label{sec:Conclusion}

If multiple vacuum-defect wormholes are present in our world,
then there must exist
a mirror world with opposite spatial orientation. The mirror world
obtained for vacuum-defect wormholes appears to be quite different from the
one discussed in the particle-physics
literature~\cite{LeeYang1956,KobzarevOkunPomeranchuk1966,%
BlinnikovKhlopov1982,SenjanovicWilczekZee1984,Linde1988,Mohapatra-etal2002,%
Das-etal2011,Berezhiani-etal-EPJC-2018,Dunsky-etal2023,%
Okun2007,Foot2014}.
(Incidentally, our vacuum-defect-wormhole spacetime
is also very different from
the one discussed in Ref.~\cite{DokuchaevEroshenko2014},
which considers a nonorientable
\emph{intra}-universe wormhole, even though that wormhole solution
is not worked out in detail.)
For this reason, we have discussed some of the phenomenology
of our hypothetical two-sheeted world.

As mentioned in Section~5 of our original paper~\cite{Klinkhamer2023}, 
the typical size $\overline{b}$ of the vacuum-defect wormholes
(assuming their relevance to Nature)
may be of the order of the Planck length,
$l_{P} \equiv (\hbar\,G/c^{3})^{1/2} \sim 10^{-35}\,\text{m}$.
In the spirit of the discussion in Section~III~H of
Ref.~\cite{MorrisThorne1988}, we can imagine that an
advanced civilization manages to ``harvest''
such a very small wormhole and then
``fattens'' it by adding a finite amount of normal matter 
(see also Appendix~B of Ref.~\cite{Klinkhamer2023}).  

But the question remains what the initial
density of these Planck-scale vacuum-defect wormholes would be.
Originally, we thought that the density would be very small,
as suggested by tight UHECR bounds on related defects in
a single spacetime (as summarized in
Appendix~\ref{app:Modified dispersion-relations-from-spacetime-defects}).
But the surprising result is that these UHECR bounds do not
apply to the vacuum-defect wormholes considered here.
Perhaps astrophysics bounds on the typical length scales of
the vacuum-defect-wormhole spacetime can be obtained by
considering scattering effects. 
Laboratory bounds can, in principle,  be obtained from
timing and apparent-brightness measurements of ultrarapid bursts
or, similar to the analysis in Sec~\ref{subsec:Imaging-bound},
from apparent-size measurements of point-like sources.

\begin{acknowledgments}
It is a pleasure to thank D.A. Muller for helpful comments 
on electron microscopy and 
the referee for thoughtful remarks on the larger physics context.
\end{acknowledgments}

\begin{appendix}
\section{First-order equations of general relativity in the vacuum}
\label{app:First-order-vacuum-equations-of-GR}

Let us briefly recall the first-order (Palatini) formulation
of general relativity, for the case that there is no matter present
(an effective classical vacuum).
The spacetime fields are then the tetrad $e^{a}_{\mu}(x)$
[building the metric tensor $g_{\mu\nu}(x)
\equiv e^{a}_{\phantom{z}\mu}(x)\,e^{b}_{\phantom{z}\nu}(x)\,\eta_{ab}$, with 
the Minkowski metric $\eta_{ab}$]
and the Lorentz connection $\omega^{\phantom{z}a}_{\mu\phantom{z}b}(x)$.

Using differential forms in the notation of Ref.~\cite{EguchiGilkeyHanson1980}
and defining the curvature 2-form
$R^{\,a}_{\,\phantom{z}b} \equiv \text{d}\omega^{\,a}_{\,\phantom{z}b}+
\omega^{a}_{\phantom{z}c}  \wedge \omega^{c}_{\phantom{z}b}$,
the first-order vacuum equations of general relativity
are~\cite{Horowitz1991}%
\bsubeqs\label{eq:first-order-eqs}
\beqa\label{eq:first-order-eqs-no-torsion}
e^{\,[\,a} \wedge D\, e^{\,b\,]} &=& 0\,,
\\[2mm]
\label{eq:first-order-eqs-Ricci-flat}
e^{\,b} \wedge R^{\,cd}\,\epsilon_{abcd} &=& 0 \,,
\eeqa
\esubeqs
with the covariant derivative
$D\, e^{b} \equiv \text{d}e^{b}+\omega^{\,b}_{\,\phantom{z}c}\wedge e^{c}$,
the completely antisymmetric symbol $\epsilon_{abcd}$,
and the square brackets around Lorentz indices denoting
antisymmetrization.
In terms of $e^{a}_{\mu}$ and $\omega^{\phantom{z}a}_{\mu\phantom{z}b}$,
these equations are manifestly
first order, as a single exterior derivative $\text{d}$
enters $D$ and $R^{\,cd}\,$.
Equation \eqref{eq:first-order-eqs-no-torsion}
corresponds to the no-torsion condition
and \eqref{eq:first-order-eqs-Ricci-flat} to the Ricci-flatness equation
[$\mathcal{R}_{\mu\nu}(x)=0$ in the standard coordinate formulation
with the Ricci tensor denoted by a calligraphic symbol, different from
the curvature 2-form $R^{\,a}_{\,\phantom{z}b}$\,].

Further discussion of the first-order formulation
of general relativity can be found in
Refs.~\cite{Schrodinger1950,Ferraris-etal-1982,Wald1984,Burton1998}.

\section{Modified dispersion relations from spacetime defects}
\label{app:Modified dispersion-relations-from-spacetime-defects}

The simplest type of static defect considered in
Ref.~\cite{BernadotteKlinkhamer2006} (called ``case-1'' there)
consists of a flat Euclidean 3-space with the
interior of a ball of radius $b_\text{def}$ removed and
antipodal points on the ball's surface identified.
For a single such defect, the topology then corresponds
to $\mathbb{R}P^{3}$, the 3-dimensional projective
space~\cite{Klinkhamer2013}.
Later, a genuine solution of the Einstein equation, with or
without matter, has been found~\cite{Klinkhamer2014,%
KlinkhamerSorba2014,Klinkhamer2013}
(see also the subsequent review~\cite{Klinkhamer2018}
with further discussion of the rather surprising phenomenology).
But for calculations of the modified dispersion relations
only the topology matters.

\begin{figure}[t]   
\vspace*{-0mm}
\begin{center}  
\includegraphics[width=0.5\textwidth]{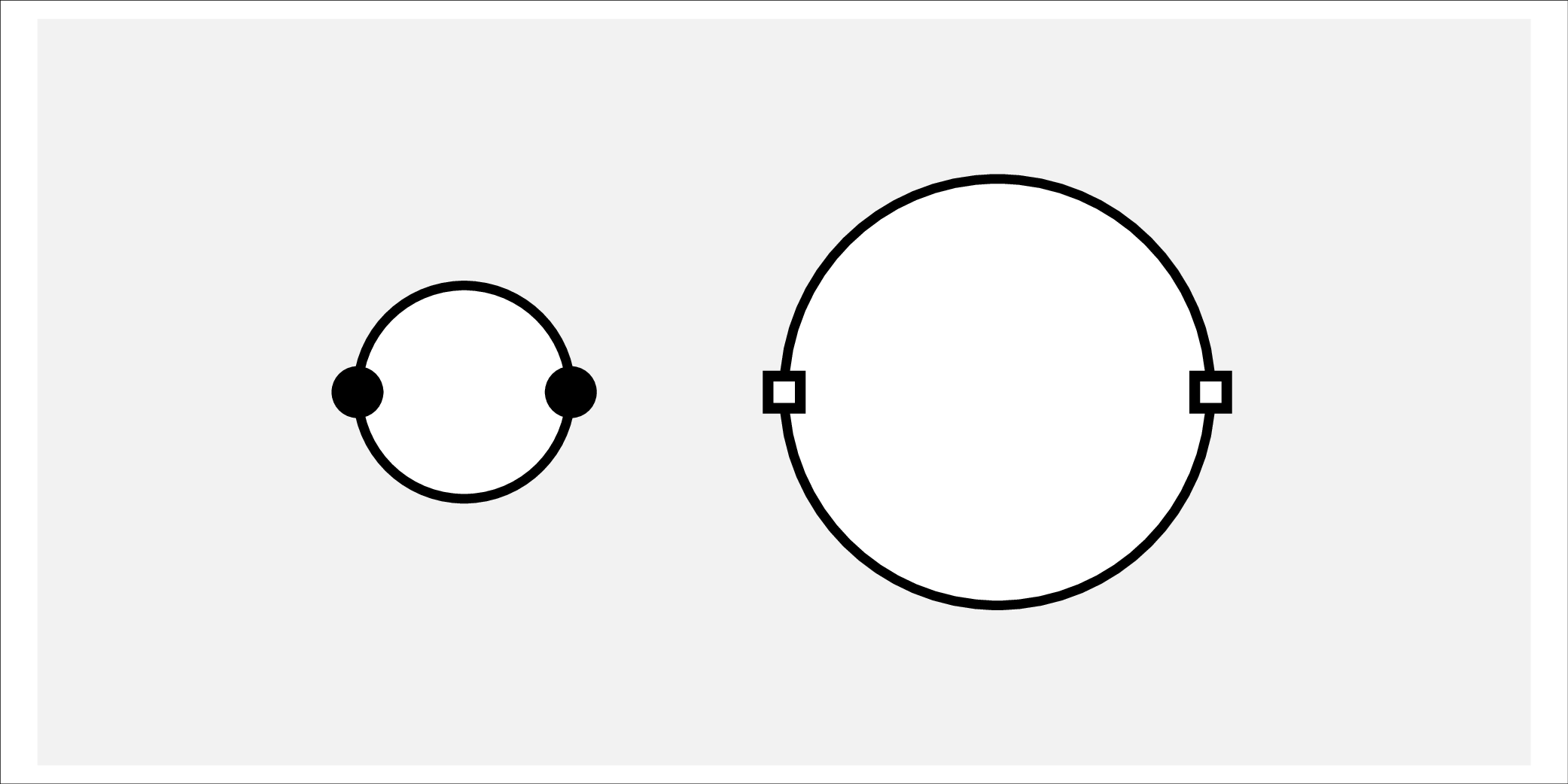}
\end{center}
\vspace*{-2mm}
\caption{Sketch of the $N=2$ multiple-defect spacetime,
at $Z_{\pm}=0$ and an arbitrary fixed time $t$.
Each defect corresponds to a 3-sphere with antipodal spacetime
points identified.
One spacetime point on the left defect is marked by a heavy dot and
a different point on the right defect by an open square.
}
\label{fig:N-is-2-multiple-defect-points}
\vspace*{00mm}
\end{figure}

Let us now consider a ``gas'' of randomly-positioned case-1 defects
(Fig.~\ref{fig:N-is-2-multiple-defect-points} shows two such defects).
Denoting the typical defect size by $\overline{b}_\text{def}$
and the typical separation between defects by $\overline{l}_\text{def}$,
the calculated photon ($\gamma$) dispersion relation
in the large-wavelength ($k\,\overline{b}_\text{def} \ll 1$)
and dilute-gas
($\overline{b}_\text{def}/\overline{l}_\text{def} \ll 1$) approximations
has the following structure
(see Eq.~(2.14) of Ref.~\cite{BernadotteKlinkhamer2006}):
\beq
\label{eq:photon-mod-disp-rel-defects}
\left[\omega_{\gamma}^\text{defect}(k)\right]^2
\sim
\left[1 - a_{2}\,
\big(\,\overline{b}_\text{def}\big/\,\overline{l}_\text{def}\big)^{3}\right]\,c^2\,k^2 
+  a_{4}\,
\big(\,\overline{b}_\text{def}\big/\,\overline{l}_\text{def}\big)^{3}\,
\overline{b}_\text{def}^{\,2}\;c^2\,k^4  + \ldots \,,
\eeq
with $k\equiv  |\vec{k}|$
and positive coefficients $a_{2} = O(10)$ and $a_{4} = O(1)$.
(Strictly speaking, the calculation of Ref.~\cite{BernadotteKlinkhamer2006}
was for a dilute gas of \emph{identical} defects, but the result
carries over to the case of having not too different defect sizes.)
The calculation in Appendix~B of Ref~\cite{BernadotteKlinkhamer2006}
showed that the dispersion relation
of a structureless Dirac particle $p$
(for example, an electron or a proton without partons)
remains unmodified to lowest order:
\beq
\label{eq:proton-mod-disp-rel-defects}
[\omega_{p}^\text{defect}(k)]^2
\sim
c^2/\lambda_{p}^2 + c^2\,k^2  +  \ldots \,,
\eeq
with the reduced Compton wavelength
$\lambda_{p}\equiv \hbar/(m_{p}\,c)$ of the particle.

From these results, we see that the photon phase velocity,
\beq
\label{eq:photon-phase-velocity}
v_{\text{ph},\,\gamma} \equiv \sqrt{(1-\kappa)/(1+\kappa)}\,c\,,
\quad \text{for}\;\;0 < \kappa \equiv
(a_{2}/2)\,\big(\,\overline{b}_\text{def}\big/\,\overline{l}_\text{def}\big)^{3}
\leq 1\,,
\eeq
is less than the maximal velocity ($c$) of the Dirac particle, so that
there can be Cherenkov radiation, but now already in
the vacuum~\cite{KlinkhamerSchreck2008}.
From the observed absence of this type of
Lorentz-violating decay processes in
ultrahigh-energy cosmic rays (UHECRs),
it is possible to obtain stringent
bounds~\cite{KlinkhamerRisse2007,KlinkhamerRisse2008,%
KlinkhamerSchreck2008}.
A recent bound gives~\cite{DuenkelNiechciolRisse2023}
\beq
\label{eq:kappa-UHECR-bound}
2\,\kappa \;\Big|^\text{(UHECR)}
< 6 \times 10^{-20}\;(98\%\,\text{CL})\,.
\eeq
With $2\,\kappa =
a_{2}\,\big(\,\overline{b}_\text{def}\big/\,\overline{l}_\text{def}\big)^{3}$
from \eqref{eq:photon-mod-disp-rel-defects}
and \eqref{eq:photon-phase-velocity}
for a Swiss-cheese-type spacetime
with static case-1 defects, we obtain
\beq
\label{eq:b-over-l-UHECR-bound}
\overline{b}_\text{def}/\overline{l}_\text{def}\;\Big|^\text{(UHECR)}
\lesssim 2 \times 10^{-7}\,,
\eeq
where we have taken $a_{2}=10$ in the $\kappa$ definition.
In other words, the defects must correspond to a
\emph{very} dilute gas, 
with the typical defect separation being more than a million times larger 
than the typical defect size.  

\end{appendix}


\end{document}